# ON DISCRETE THERMODYNAMICS OF ELECTROCHEMICAL SYSTEMS AND ELECTROCHEMICAL OSCILLATIONS

B. Zilbergleyt[1]

ABSTRACT

The article presents results of discrete thermodynamics (DTD) basic application to electrochemical systems. Consistent treatment of the electrochemical system as comprising two interacting subsystems - the chemical and the electrical (electrochemical) - leads to ln-logistic map of states of the electrochemical system with non-unity coefficient of the electrical charge transfer. This factor provides for a feedback and causes dynamic behavior of electrochemical systems, including bifurcations and electrochemical oscillations. The latter occur beyond bifurcation point at essential deviation of the chemical subsystem from true thermodynamic equilibrium. If the charge transfer coefficient takes on unity, the map turns into classical equation of electrochemical equilibrium. Spectra of electrochemical oscillations, resulted from the DTD formalism, are multifractals. Graphical solutions of this work are qualitatively compared to some experimental results.

INTRODUCTION

Although classical and irreversible thermodynamics of electrochemical systems are both considered well established parts of the electrochemical science, no sign of electrochemical oscillations can be drawn out of them; all known up to now models of any electrochemical (and just chemical as well) oscillations are based exclusively on chemical/electrode kinetics. As opposite to that, in papers on discrete thermodynamics of chemical equilibria (DTD), recently published by the author, thermodynamic instabilities, occurring under impact of external thermodynamic force (TdF) were shown to be the immediate source of possible oscillations in closed chemical systems. First time we encountered them developing discrete thermodynamics of lasers [1], then more information was added in [2].

True thermodynamic equilibrium (TdE) is defined only for isolated systems. According to Le Chatelier principle, being shifted from TdE by an external force, thus becoming either closed or open, the system tends to a new state of equilibrium, not identical to TdE by virtue of the system non-isolation; in chemical systems it matches the state of *chemical equilibrium*. Basic expressions, allowing us to find distance between TdE and the shifted equilibrium state in terms of reaction extent – ln-logistic maps of state, were derived in our previous works in supposition that the system tries to achieve the state with balanced internal (bound affinity [3]) and external thermodynamic forces. This paper offers a new thermodynamic approach to electrochemical systems and electrochemical oscillations, based on ideas of discrete thermodynamics. The reader is advised to take a look at [2] to refresh the DTD basics before reading further on.

EQUILIBRIUM OF ELECTROCHEMICAL CELL

We think of chemical system dynamics in terms of subsystems, interacting by means of thermodynamic forces and subsystem deviations from thermodynamic equilibria in terms of their process extents (or process coordinates). This model may be applied with essential gain to many, not only chemical systems, e.g. lasers [1]. In such an approach, the electrochemical cell comprises two subsystems – the chemical and the electrochemical (actually, electrical); they are inseparable and functionally tied, their states change strongly interdependently. In conventional terminology both are the closed systems, there is no material exchange between them. Electrochemical cell equilibrium means equilibrium between them, when driving thermodynamic forces are mutually balanced: at the cell equilibrium electrode potential supports *chemical shift* from TdE, being at the same time caused by it. If the electrical current in a closed circuit is slow, the states of subsystems change simultaneously, and the cell still stays in electrochemical equilibrium. Various reasons, usually summarized in overpotential, make subsystem changes asynchronous and irreversible, causing delays on the current - voltage graphs.

---

[1] System Dynamics Research Foundation, Chicago, USA, e-mail: sdrf@ameritech.net.

Force is defined in physics as negative derivative of a potential by coordinate. TdF in chemical thermodynamics, or thermodynamic affinity was introduced by De Donder as a negative derivative of Gibbs' free energy by the reaction extent [4]. De Donder measured reaction extent in moles, and his affinity has dimension of kJ/mol$^2$, a hard to explain result. Discrete thermodynamics defines affinity $A_{jc}$ at p,T=const in finite differences as negative ratio of the j-system Gibbs' free energy change $\Delta G_j$ to *dimensionless* reaction extent increment $\Delta\xi_{jc}$

(1) $\qquad\qquad\qquad\qquad\qquad\qquad\qquad\qquad\qquad\qquad A_{jc} = -\Delta G_j/\Delta\xi_{jc}$,

subscript "c" relates values to the chemical subsystem; shift of the latter from TdE is

(2) $\qquad\qquad\qquad\qquad\qquad\qquad\qquad\qquad\qquad\qquad \delta\xi_{jc} = 1 - \Delta\xi_{jc}$,

($\Delta_{jc}$ and $\delta_{jc}$ further on in writing). Affinity in DTD has dimension of energy. Definition of the electrical thermodynamic moving force is not as simple as chemical, and usual for non-equilibrium thermodynamics acceptance of the $z_iE$ value, a product of the ion charge by the electrode potential, as TdF [5] is definitely oversimplified. Electrical TdF cannot be derived from general definition of the thermodynamic force because it seems to do not exist at all, and the term "generalized force" is quite often to hide this. Expression for appropriate TdF depends upon nature of the model of question and its thermodynamic description. To avoid some relevant clumsiness, in our previous publications external TdF, acting against a chemical or a quasi-chemical system, was replaced by power series of the system shift from TdE forced by that force. Now we are going to use thermodynamic forces explicitly, and we have to define appropriate force and the process coordinate for the electrical subsystem.

Spontaneous move of electrochemical system towards equilibrium follows the $E_j \rightarrow E^{eq}_j$ direction, where $E_j$ and $E^{eq}_j$ are running and equilibrium electrode potentials; the extent of electrochemical process may be defined as

(3) $\qquad\qquad\qquad\qquad\qquad\qquad\qquad\qquad\qquad\qquad \Delta_{je} = E_j/E^{eq}_j$,

corresponding shift of the electrochemical subsystem from equilibrium is

(4) $\qquad\qquad\qquad\qquad\qquad\qquad\qquad\qquad\qquad\qquad \delta_{je} = 1 - \Delta_{je} = 1 - E_j/E_j^{eq}$;

Because $E_j = E_j^{eq} + \chi_j$, where $\chi_j$ is overpotential[1], $\Delta_{je} = (E_j^{eq} + \chi_j)/E_j^{eq} = 1 + \chi_j/E_j^{eq}$, and we arrive at the shift expression as

(5) $\qquad\qquad\qquad\qquad\qquad\qquad\qquad\qquad\qquad\qquad \delta_{je} = -\chi_j/E^{eq}_j$.

The shift shows how far the subsystem falls apart from electrochemical equilibrium, where $E_j = E^{eq}_j$, $\Delta_{je} = 1$ and $\delta_{je} = 0$. Possible inequalities $E_j < E^{eq}_j$ and $E_j > E^{eq}_j$ obviously lead to $\Delta_{je} < 1$, $\delta_{je} > 0$ and $\Delta_{je} > 1$, $\delta_{je} < 0$. Because overpotential originates not from thermodynamic reasons and its value is often unpredictable, *the chemical subsystem shift in general is not identical to the electrochemical shift*.

Now, the electrochemical thermodynamic force may be expressed via the equivalent of the cell electrical energy change as cell potential changes from zero to $E_j$, or $n_j\mathcal{F}E_j$, divided by the electrochemical system process extent $\Delta_{je}$, or

(6) $\qquad\qquad\qquad\qquad\qquad\qquad\qquad\qquad\qquad\qquad TdF_{je} = n_j\mathcal{F}E_j/\Delta_{je}$,

$n_j$ is the amount of transferred electron charges, $\mathcal{F}$ – Faraday number as usually, and $E_j = E_j^{eq} + \chi_j$. The electrochemical cell equilibrium corresponds to the balance of thermodynamic forces, driving the subsystems

(7) $\qquad\qquad\qquad\qquad\qquad\qquad\qquad\qquad\qquad\qquad \Delta G_j/\Delta_{jc} = n_j\mathcal{F}E_j/\Delta_{je}$.

Recalling (2) and (3) one can obtain

---

[1] Traditional for electrochemistry Greek letter η for overpotential was taken by DTD to mark thermodynamic equivalent of transformation [2].

(8) $$\Delta G_j - (1-\delta'_{jc})n_j\mathcal{F}E_j^{eq} = 0,$$

tick mark at $\delta'_{jc}$ refers to the chemical subsystem shift from TdE at the electrochemical equilibrium. *Each cell has its own characteristic equilibrium electrode potential which is in agreement with a certain shift of the chemical subsystem from TdE, denoted by tick mark.* Let us reiterate: map (8) describes the electrochemical cell equilibrium when its chemical subsystem is not at TdE ($\delta'_{jc} \neq 0$!), forces that move subsystems are equal – the cell potential is balanced by the changes in the system chemical/ionic composition. At $\delta'_{jc} = 0$ map (8) describes exactly TdE of the chemical subsystem, turning into classical equation of electrochemical equilibrium

(9) $$\Delta G_j - n_j\mathcal{F}E_j^{eq} = 0.$$

Here is a contradiction: at the state, described by (9), the chemical subsystem must be isolated from its environment *by definition* and has no relation to its electrical counterpart – now the electrochemical system, comprising interacting subsystems, does not exist any more as *a system*, and such electrochemical equilibrium is not possible. Map (8) may be used to calculate equilibrium cell potential at TdE and within the area of reversible changes as well. Indeed, taking into account that

(10) $$\Delta G_j = -RT\ln K_j + RT\ln[\Pi(a_{ij})]$$

and presenting logarithmic members as functions of $\eta_j$ (thermodynamic equivalent of transformation) and $\delta_j$ (equilibrium constant corresponds to $\delta'_{jc}=0$) [2], neglecting non-ideality we arrive finally at a map, ready to simulate electrochemical equilibrium

(11) $$\ln[\Pi(\eta_j,0_j)/\Pi(\eta_j,\delta'_{jc})] + (1-\delta'_{jc})n_j\mathcal{F}E_j^{eq}/RT = 0.$$

This map reflects the fact that if the chemical system hasn't achieved its TdE, the electron transfer coefficient is less than unity and equals to $(1-\delta'_{jc})$[I]. Nernst equation, following from the DTD, turns into a map

(12) $$(1-\delta'_{jc})E_j^{eq} = E_j^0 - (RT/n_j\mathcal{F})\ln\Pi(\eta_j,\delta'_{jc})],$$

showing explicitly that the shift related factor at the equilibrium potential has roots in the chemical subsystem deviation from TdE.

ELECTROCHEMICAL SYSTEM DOMAIN OF STATES AND OSCILLATIONS

Factor $(1-\delta_j)$ in the above derived maps provides a feedback, turning the electrochemical cell into a dynamic system; graphical solutions to this map are specific fork bifurcation diagrams. Fig.1 shows combined simulation results in $\delta_{jc}$ vs. $E^{eq}_j$ coordinates – one of map (11) and another one, quasi-classical simulation of map (9) (named "DTD" and "classical" respectively in the picture). The latter followed the pattern, established earlier [2]: because notion of the shift is strange for classical thermodynamics, a conditional shift value was included in simulation – system equilibrium state at $E^{eq}_j = 0$ with $\Delta_j=1$, $\delta_j=0$ was appointed as TdE, while all other states (which are all TdE in classical theory) were accounted for as equilibria, distant from the reference state by individual shifts. At a glance, both curves in Fig.1 were obtained within similar ideologies: the base state matching TdE and all other were considered as shifted from the base. However, the principal difference is that the DTD map derivation included thermodynamic forces explicitly while the "classical" is based on minimization of the system free energy. Fig.1 reveals the difference between the DTD and the classical behavior of electrochemical systems.

Bifurcation diagrams of map (11) do not follow a multiple period doubling (so called Feigenbaum route, [6]): instead, they experience only one time bifurcation period 2, then one can see instabilities that reveal themselves as well pronounced oscillations between the upper and the lower bifurcation branches as the external force ($\Delta\varphi$) changes. The oscillation spectra are linear, located within a restricted interval of the external TdF, and line or the line groups are separated by clearly visible windows of stability. We found that every set of the map (11) parameters has its own signature in oscillation spectra. Changing parameters in map (11) one can create a full set of bifurcation diagrams which constitute the electrochemical system *domain of states*, totally covering the coordinate plane ($\delta_j$ and $\Delta\varphi_j$ both may take on plus and minus signs independently). Electrochemical oscillations exist within the area, restricted by certain values of the reaction standard change of Gibbs' free energy and certain values of potential difference on the cell. These features for a simple red/ox reaction A+B=AB (say, A−ne⁻=$A^{n+}$, B+ne⁻=$B^{n-}$) are exemplified in Fig.2 and Fig.3. Oscillations occur within the cusped bodies.

---

[I] This can be used for interesting development of the Batler-Volmer equation with new meaning of the charge transfer coefficients. One can easily prove in the DTD formalism that cathode and anode shifts are related as $\delta_{jcat} = 1-\delta_{jan}$, thus providing for traditional relationship between the transfer coefficients.

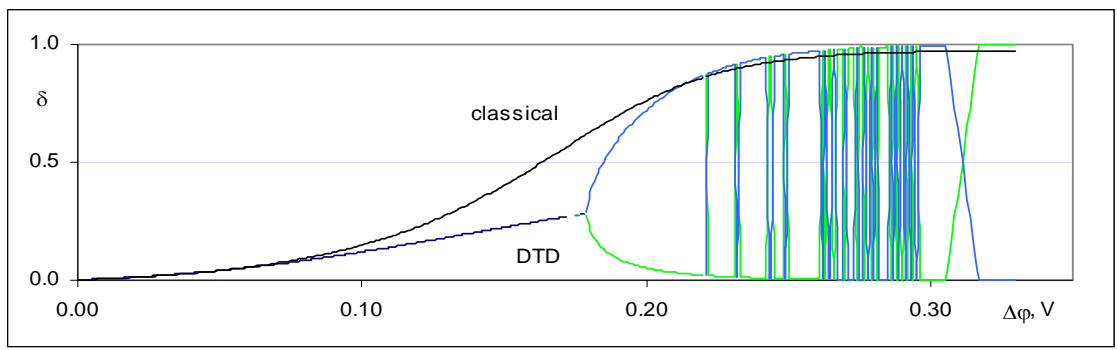

Fig.1. Comparison btw "classical" and DTD shift dependence upon equilibrium cell potential. Redox reaction A+B=AB, $n_j$=1, $\Delta G^0_j$=−18.0 kJ/m, T=293K.

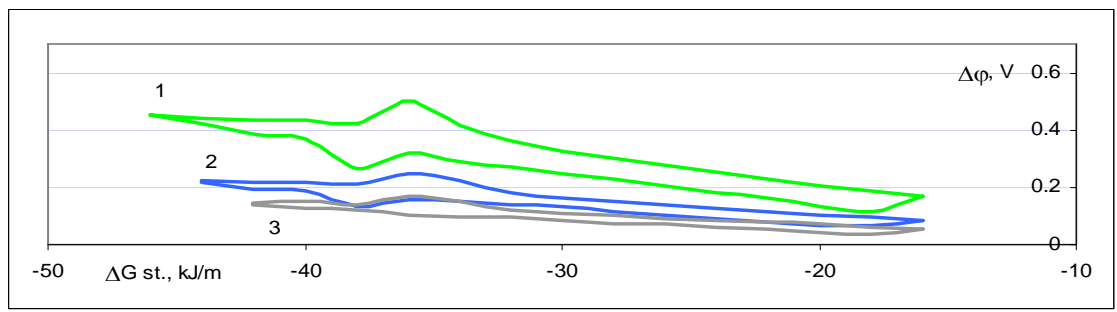

Fig.2. Electrochemical oscillations areas for different amount of transferred electrons (numbers at the curves), reaction A+B=AB, T=293.15K.

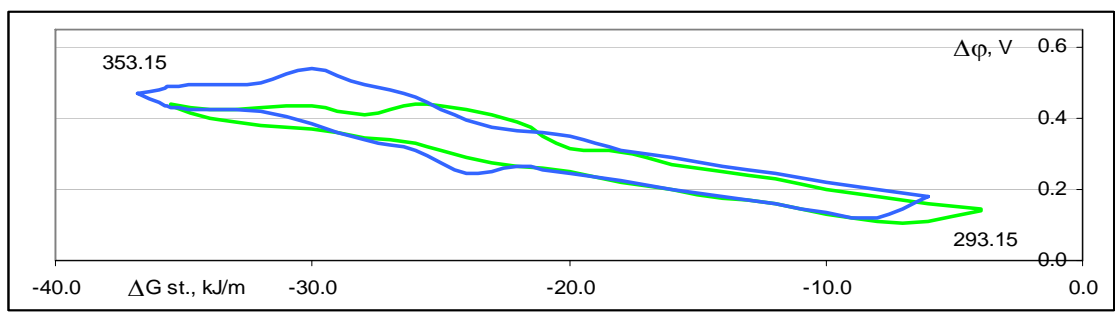

Fig.3. Electrochemical oscillations areas for different temperatures (shown at the curves), reaction A+B=AB, temperature were varied separately from $\Delta G^0$, $n_j$=1.

Another feature of the found oscillations is their fractal nature. To prove it, the task we set was identical to the length of the shore line [7]. In simulation, the $\Delta\varphi$ step was varied; the results are shown in Fig.4. One can see that the oscillation spectra are multifractals indeed.

Map (11) is restricted by electrochemical equilibrium and actually describes equilibrium steady state; what happens when the circuit gets closed? Electrical current leads to changes in the cell potential; the electrical subsystem takes a lead over its chemical counterpart, prompting it to change. The same happens in case of electrolytic cells. If changes occur in a reversible, slow pace, both subsystems follow closely each other in concord, still obeying map (11). When the changes speed up beyond the reversibility limit, the subsystems fall out of concert, the chemical subsystem is behind, and overpotential is added to the cell potential. Overpotential originates from non-thermodynamic reasons, and cannot be correctly incorporated into the body of DTD model of electrochemical equilibrium.

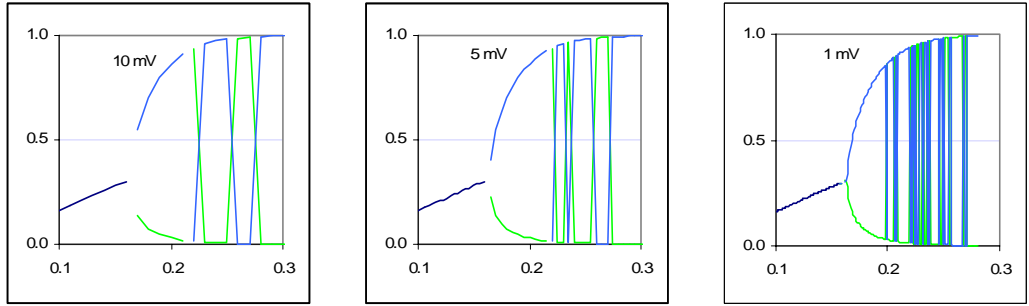

Fig.4. Fractal properties of the electrochemical oscillation spectra, $\delta_j$ (ordinate) vs. $\Delta\varphi$, V (abscissa), "electrical stick" lengths are shown on the pictures.

COMPARISON TO ALTERNATIVE DTD MAP OF STATES

An alternative and more general way to obtain the system map of states, already mentioned, was developed in our previous works [2]. It consists in presenting all thermodynamic forces in terms of the system shift from TdE; applying the same to the electrochemical systems with low complexity, we arrive at a regular chemical system states map

(13) $\qquad \ln[\Pi_j(\eta_j,0)/\Pi_j(\eta_j,\delta_j)] - \tau_j\delta_j(1-\delta_j)=0$,

where $\tau_j$ is a factor, defining growth of the system deviation from TdE. Through division by $(1-\delta_j)$ turns map (13) into a balance of thermodynamic forces acting against j-subsystem; corresponding graphical solution in coordinates $\delta_j$ vs. external TdF, or dynamic bifurcation diagram is shown in Fig.5. General similarity between diagrams for electrochemical and chemical systems is obvious, and both feature a trend of the spectra lines grouping into packages. Also, in both cases the spectra, obtained by iterations of maps (11) and (13), are fractals, whose characteristics depend on the iteration steps. For example, for the multifractal oscillations [map (13)] related to reaction $PCl_3+Cl_2=PCl_5$ ($\Delta G^0=-26.93$ kJ/m at T=348.15K) with varying "chemical stick", i.e. iteration step for the shift $\delta_j$, the Hausdorff dimension [8] limit at zero "stick" length is $\approx 0.94$ (to compare with other fractals see [9]). In both cases oscillations occur within certain intervals of the standard change of Gibbs' free energy. The advantage of map (11) is its closeness to concrete objects while map (13) is more general, allowing for more detailed look at the chemical system behavior.

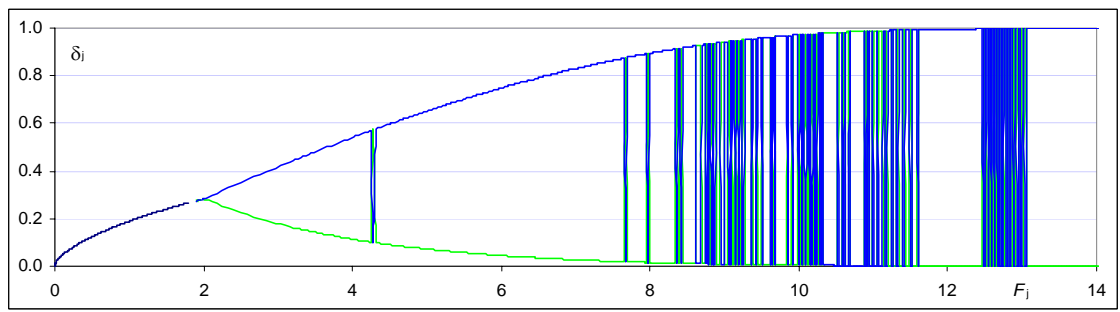

Fig.5. Chemical system oscillations, shift vs. external thermodynamic force,
A+B=AB, $\Delta G^0_j = -18.0$ kJ/m, T=293K.

COMPARISON TO EXPERIMENTAL DATA

Our results were plotted in $\delta_j$–E curves while experimental points are usually presented in *I*-E coordinates. There must exist a correlation between oscillations of the electrical current and the chemical subsystem states in the electrochemical systems, where both subsystems closely follow each other. We don't know it yet, and we looked only for a qualitative similarity between our results and experimental data. It should be mentioned that experimental results often reveal only one branch, making the experimental graphs similar to a half graph from Fig.1 (Fig.6).

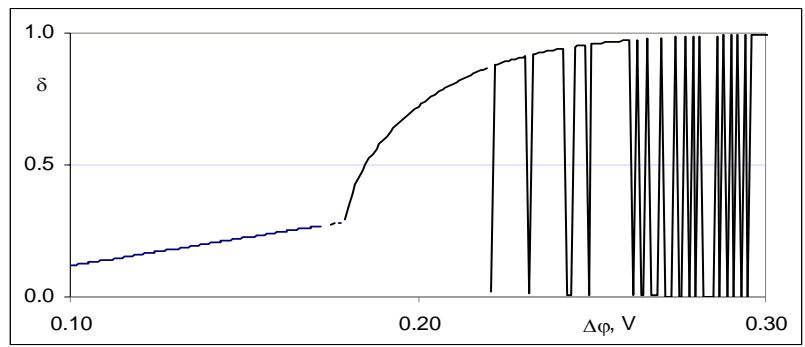

Fig.6. A part of the Fig.1 half-graph, A+B=AB, the upper branch.

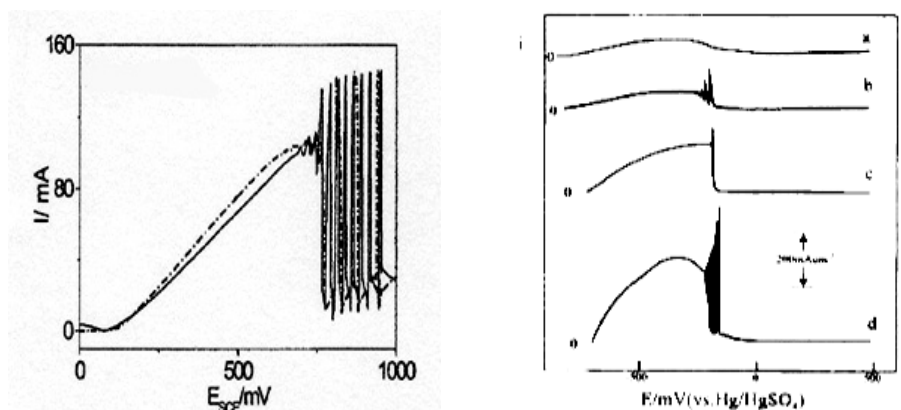

Fig.7, left. Polarization behavior of the $Cu/CCl_3COOH$ system; concentration of $CCl_3COOH$ – 1 M/dm3, adopted from [10].

Fig.7, right. Electrochemical oscillations in $Fe/HNO_3$ system at different concentrations of nitric acid, adopted from [11].

Experimental graphs, adopted from various publications are shown in Fig.7 and Fig.8. The authors of [10] have mentioned that the origin of the oscillatory behavior of metals in acidic electrolyte (Fig.7, left) was not well understood and they supposed that the oscillations were caused by periodic film formation-dissolution on the electrode surface as well as periodic changes of the film chemical contents. Admitting purely kinetic reasons, the authors didn't admit that the film transformations may result from changes of the chemical subsystem states, i.e. might be caused thermodynamically. As well as the system states were definitely dependent on the nitric acid concentrations in [11] (Fig.7, right). It was found in [10] that the oscillations occurred within the potential range ≤0.3 V for all investigated concentrations of the trichloroacetic acid solutions. One can find more examples of similar coincidences. Also, we have reproduced in Fig.8, right, a picture of potential oscillations in open circuit [13]. The inset is the magnification; one can see that in some examples the amplitudes of singular oscillation cycles are restricted as if by the bifurcation branches. This is the case when one can suppose with certain probability the leading role of chemical subsystem.
Following similarities between the thermodynamically predicted and experimentally observed electrochemical oscillations may be noticed:
- oscillation areas are located within relatively narrow voltage interval;
- oscillation spectra experience windows within which the system is undisturbed;
- in many cases experimental oscillations magnitudes are restricted by lines similar to lower and upper bifurcation branches.

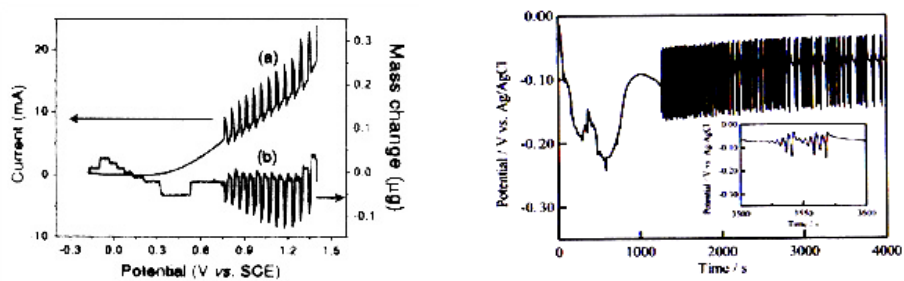

Fig.8, left. Linear sweep voltammogram at a platinized Pt electrode, 0.1 M $H_2SO_4$+1 M HCHO, a) current-potential plot, b) EQCM mass change plot, adopted from [12].
Fig.8, right. Open circuit potential oscillations on porous silicon during immense precipitation of Cu from 0.01 M $CuSO_4$+0.1% HF under stagnant condition, adopted from [13].

CONCLUSION

This paper does not offer systematical application of discrete thermodynamics to electrochemical equilibria; we just have investigated and described only a couple of key points were the DTD results differed from the classical ones. To the best of the author's knowledge, the electrochemical oscillations, that are *thermodynamically* predictable, were never described before. One should mention that although no one of the found oscillation similarities could be directly related to changing states of the chemical subsystem, the explanations given by the authors of the papers with observed oscillations were also just suppositions, leaving enough space for some other hypotheses and speculations.
It is well recognized that results of thermodynamic analysis depend upon thermodynamic description of the system. Derived in this work ln-logistic map of states for electrochemical systems explicitly shows relationship between the chemical subsystem shift from thermodynamic equilibrium and electrode potential, reveals instabilities of the bifurcation branches and appropriate oscillations of the chemical subsystem states. Similarly to what was found in all other previously investigated cases, the system experience onset of instabilities and oscillations when it essentially deviates from TdE (it might be said that it reaches to the "far-from-equilibrium" area). When those oscillations happen to electrochemical system, there is no other way for relaxation than via electrical current; this leads to pictures, similar to experimentally observed current oscillations vs. electrode potential.

In kinetic explorations of electrochemical oscillations, the chemical gross-reaction is usually split by steps, each consisting of two connected by arrow stages to show the transformation directions and differential equations. Joint solution to the system of such equations under certain, *ad hoc* chosen conditions and *ad hoc* numerical (quasi-stoichiometric) coefficients leads to chemical oscillations, as a rule occurring within a certain restrictions of the equations parameters. This is how series of well known models like "brusselator", "oregonator", etc. were born. Also, a statement of chemical instabilities as a reason of chemical oscillations is usually attached to most kinetic oscillatory models, though quite often there were no explicit accounts for any instability in most of them at all. Title of one publication– "Chemical Oscillations Arise Solely from Kinetic Nonlinearity and Hence Can Occur Near Equilibrium" [13] – may serve as a manifesto for such an approach. Kinetic perception of chemical oscillations reigns in numerous publications and all basic monographs like [14, 15] and many others. We have no doubts that kinetic models are able to produce images, close enough to experimentally observed chemical oscillations. This paper offers alternative explanation and questions the exclusivity of kinetic approach.